\documentclass[12pt]{article}
\usepackage{a4,graphicx}
\usepackage{amsmath}
\usepackage{amssymb}
\usepackage{mathbbol}
\renewcommand{\theequation}{\thesection.\arabic{equation}}

\begin{document}
\title{Localisation of fermions to brane: Codimension $d\ge 2$}
\author{{\large W. Nahm}$^{\star}$,
and {\large D. H. Tchrakian}$^{\dagger \star}$ \\ \\
$^{\star}${\small School of Theoretical Physics, Dublin Institute for
Advanced Studies}\\
{\small 10 Burlington Road, Dublin 4, Ireland}\\ \\
$^{\dagger}${\small Department of
Mathematical Physics, National University of Ireland Maynooth,} \\
{\small Maynooth, Ireland}}
\newcommand{\dd}{\mbox{d}}
\newcommand{\tr}{\mbox{tr}}
\newcommand{\la}{\lambda}
\newcommand{\ka}{\kappa}
\newcommand{\al}{\alpha}
\newcommand{\ga}{\gamma}
\newcommand{\de}{\delta}
\newcommand{\si}{\sigma}
\newcommand{\bomega}{\mbox{\boldmath $\omega$}}
\newcommand{\bsi}{\mbox{\boldmath $\sigma$}}
\newcommand{\bchi}{\mbox{\boldmath $\chi$}}
\newcommand{\bal}{\mbox{\boldmath $\alpha$}}
\newcommand{\bpsi}{\mbox{\boldmath $\psi$}}
\newcommand{\brho}{\mbox{\boldmath $\si$}}
\newcommand{\beps}{\mbox{\boldmath $\varepsilon$}}
\newcommand{\bxi}{\mbox{\boldmath $\xi$}}
\newcommand{\bbeta}{\mbox{\boldmath $\beta$}}
\newcommand{\ee}{\end{equation}}
\newcommand{\eea}{\end{eqnarray}}
\newcommand{\be}{\begin{equation}}
\newcommand{\bea}{\begin{eqnarray}}
\newcommand{\ii}{\mbox{i}}
\newcommand{\e}{\mbox{e}}
\newcommand{\pa}{\partial}
\newcommand{\Om}{\Omega}
\newcommand{\vep}{\varepsilon}
\newcommand{\bfph}{{\bf \phi}}
\newcommand{\lm}{\lambda}
\def\theequation{\arabic{equation}}
\renewcommand{\thefootnote}{\fnsymbol{footnote}}
\newcommand{\re}[1]{(\ref{#1})}
\newcommand{\R}{{\rm I \hspace{-0.52ex} R}}
\newcommand{\N}{{\sf N\hspace*{-1.0ex}\rule{0.15ex}%
{1.3ex}\hspace*{1.0ex}}}
\newcommand{\Q}{{\sf Q\hspace*{-1.1ex}\rule{0.15ex}%
{1.5ex}\hspace*{1.1ex}}}
\newcommand{\C}{{\sf C\hspace*{-0.9ex}\rule{0.15ex}%
{1.3ex}\hspace*{0.9ex}}}
\newcommand{\eins}{1\hspace{-0.56ex}{\rm I}}
\renewcommand{\thefootnote}{\arabic{footnote}}

\maketitle


\bigskip

\begin{abstract}
We investigate
$4+d$ dimensional fermionic models in which the system in codimension-$d$
supports a topologically stable solution, and in which the fermion may be
localised to the brane, with power law in 'instanton' backgrounds and
exponentially in 'soliton' backgrounds.
When the fermions are isoscalars, the mechanism fails, while for isospinor
fermions it is successful. As backgrounds we consider instantons
of Yang--Mills and sigma models in even codimensions, solitons
of sigma models in odd codimensions, as well as solitons of Higgs
and Goldstone models in all codimensions.
\end{abstract}
\medskip 
\medskip
\newpage

\section{Introduction}
The idea that our world is a (fairly flat) 3-brane in a higher dimensional
space has deep roots in the 19th century \cite{A}. It was introduced into
physics in the context of cosmological defects on one hand and the branes
of M-theory on the other. In any case one has to explain why the observed
fermion masses are so much smaller than the mass scale given by the
transversal extension of the brane. On some parameter range one expects the 
branes to approach classical configurations, such that a semiclassical
description of matter on the brane is possible. Qualitative features like
the appearance of low energy excitations on the brane should be independent
of the parameters at least locally, such that these semiclassical descriptions
may be relevant for realistic cases.

We study the extension of the original work of Rubakov and
Shaposhnikov~\cite{RS} localising a fermion to the brane, in which
a 5 dimensional model, i.e. one with codimension-$1$ is employed.
There~\cite{RS,R}, the brane Lagrangian is the $\phi^4$ (or sine-Gordon)
system in $1$ dimension, supporting the kink-soliton. We are
concerned with the possibilty of extending this mechanism to the
case of arbitrary codimension-$d\ge 2$. In general, fermions are
localised when their reduced Dirac equation in the $d$ dimensions
transversal to the brane has zero modes. This always is the case when
the index of the corresponding Dirac operator is positive. 

In case~\cite{RS} the wave function of the fermion drops exponentially
away from the brane. Such backgrounds will be called solitonic. For them 
an explicit formula for the index
is known \cite{BS}, and physically relevant specialisations are known both
in odd and in even dimension \cite{Ca, EW}. The index only depends on the
behaviour of the scalar fields at infinity, gauge fields are assumed to
approach zero at infinity and do not influence the index in a direct way. 
Nevertheless, the corresponding internal symmetries are crucial.
In the backgrounds of Yang-Mills or sigma model instantons, the fermionic
fields have a power-law fall-off away from the brane. In such cases
the available index calculations depend on compactifications at infinity,
which allow an application of the Atiyah-Singer index theorem \cite{AS}.

For solitonic backgrounds, non-zero indices occur in both even and odd
dimensions, whereas in the instanton case one needs even dimensions.
Even when the index is zero or negative, solutions of the reduced Dirac
equation might exist. For fermions in instanton backgrounds this can
be excluded by vanishing theorems for the kernel of the corresponding
Dirac operator (the Weizenb\"ock formula), but in other contexts little is
known. In particular, no general results for solitonic backgrounds seem
to exist.

The index formulas require cumbersome if straightforward integrations, but
often an analytic approach is possible. When a background can be deformed 
into a superposition of well separated configurations with spherical symmetry,
the index is additive and only these spherically symmetric situations
have to be considered. For any of them the index is the sum over the
indices for fixed angular momentum, which are determined by the behaviour
of ordinary differential equations and easy to calculate. We consider
some families of such configurations. As new features we obtain that 
certain solutions are universal and apply to any number of transversal
dimensions, and in certain cases a vanishing theorem can be established.

We have considered two alternative types of fermionic models in $4+d$
dimensions. In the first of these, ({\it i}), the Dirac spinor is
{\em isoscalar}, such that the Dirac operator on the codimension involves
only unmodified partial derivatives and no gauge field. In the second
type, ({\it ii}), the Dirac spinor on the codimension is in general an 
{\em isospinor} under an internal $SO(d)$ group, i.e. it is a square
matrix valued array. The Dirac operator features a covariant derivative,
but the index is determined for the case of a vanishing gauge field.
Here we also consider an exceptional $d=2$ case when the covariant
derivative is Abelian and the Dirac spinor is a two-component column.

Like in \cite{RS,R}, our brane Lagrangians in $d$ dimensions support
finite energy topologically stable solutions, both  of 'instanton' and of
'soliton' types. Our nomenclature here is the following: Consider the
topological charge density $\varrho[\varphi]$, e.g. the Chern--Pontryagin
(C-P) density and its descendents when $[\varphi]$ symbolises YM fields
(in even dimensions) and YM-Higgs fields, respectively, or, the degree of
the map when $[\varphi]$ stands for sigma model fields. The scalar
deansity $\varrho[\varphi]$ is {\em essentially total divergence}, in the
sense that its variation vanishes when it is subjected to arbitrary
variations $\delta\varphi$. (Indeed in the
(C-P) case $\varrho[\varphi]$ is precisely a {\em total divergence}
$\varrho[\varphi]=\pa_i\Omega_i$, $\Omega_i$ being the Chern-Simons
density.) When $\varrho[\varphi]$ is subjected to spherical symmetry,
which is the case for the fields $[\varphi]$ at infinity, it reduces
always to a {\em total derivative} of some function $\si$, i.e.
\be
\label{sigma}
\int\varrho\ d^dx\propto\int\frac{d\si}{dr}dr\,.
\ee
We denote this function pertaining to the classical
solutions $\varphi_c$ by $\si_c=\si[\varphi_c]$.
The profiles \re{inst} and \re{soli} stated below in section {\bf 2.1}
specify our nomenclature of 'instanton' and 'soliton' types, respectively.
Thus, finite energy topolologically stable solutions to Yang-Mills (YM) or
a sigma model systems in even $d$ are typified by 'instanton' profiles,
\re{inst}, and solutions to sigma models in odd $d$, or to Higgs models,
i.e. YM-Higgs (YMH) systems, or Goldstone models, in both even and odd
$d$, by 'soliton' profiles \re{soli}. By
Goldstone model, in turn, we mean the gauge decoupled version of a Higgs
model, provided of course that the soliton persists in the gauge
decoupling limit of the YMH model in question. In this context, the
solutions of the symmetry breaking model(s) employed in \cite{RS,R} are
typified by 'soliton', \re{soli}, profiles.

The main feature of the mechanism of \cite{RS,R} is dimensional descent
of a $4+d$ dimensional fermionic model to $4$ dimensional Minkowski space,
the Dirac field on which is assumed to be chiral and to be massless.
For codimension-$d\ge 2$, after the descent one is left with a nontrivial
Dirac equation  on the codimension-$d$, which we have called  the
{\em residual Dirac equation}. The crucial step is that of finding
normalisable zero modes of this residual Dirac equation. The asymptotic
behaviour of the residual Dirac spinor is then responsible for the decay
of the Dirac field off the brane. Whether this decay is achieved, and
if so, is it power like or exponential, will
depend on the $4+d$ dimensional fermionic model chosen.

It will turn out that the desired normalisable zero modes do not exist
for models of type ({\it i}) for any codimension $d$, and that such
solutions exist for models of type ({\it ii}) for all $d$.

To date, extension to higher codimensions ($d\ge 2$) for this mechanism
has been performed for codimension-$2$, in a series of works started
by Libanov and Troitsky~\cite{LT,LTeco} in flat space, and in the  presence
of gravity, in \cite{RjS}. The models employed in both \cite{LT} and
\cite{RjS} are of the type ({\it ii}), namely featuring a covariant
derivative in the Dirac operator. In these models~\cite{LT,RjS}, what
guarantees the existence of zero modes of the residual Dirac equation
in codimension-$2$, is the choice of {\it such} a Yukawa coupling in the
$6$ dimensional model {\it that} leads to a residual Dirac equation which
coincides with the particular Dirac equation on an Abelian-Higgs
background for which Jackiw and Rossi~\cite{JR} have constructed the zero
modes explicitly. To follow this line of approach for
codimensions-$d\ge 3$, one is naturally led to considering such $4+d$
dimensional fermionic models which result in $d$ dimensional residual
Dirac equations for which we know there exist normalisable zero modes, or
better still that we can construct such solutions. Dirac equations in $d$
dimensions in the background of a YM 'instanton' or a YMH 'soliton' (or
its associated Goldstone 'soliton') supporting such zero modes are the
natural candidates which will be proposed. Dirac equations in $d$
dimensions will be solved in the appropriate 'instanton' or 'soliton'
backgrounds, supported respectively by a hierarchy of YM and YMH models in
these dimensions.

The first in the hierarchy of YM models is the usual
YM system in $4$ dimensions, and its extensions to all even dimensions
as given in \cite{YM} support 'instantons' in these dimensions,
analogous to the $4$ dimensional instanton~\cite{BPST}. It is these
'instantons' in $d=2n$ dimensions that we will employ as backgrounds
for the construction of the zero modes of the residual Dirac equations,
extending\footnote{We restrict our considerations to Dirac equations on
spherically symmetric backgrounds only and exclude the multicentre
backgrounds employed in \cite{JR4}, or even periodic backgrounds
used in \cite{Chak}.} the $d=4$ result of Jackiw and Rebbi~\cite{JR4} to
arbitrary $d$.

The first in the hierarchy of non Abelian YMH models is the Georgi-Glashow
(Higgs) model in $d=3$, which supports the the usual monopole~\cite{mono}.
The zero modes  of the Dirac equation on this background was given long
ago by Jackiw and Rebbi~\cite{JRb}. The
extension to arbitrary $d$ is completely straightforward: The result of
\cite{JRb} follows directly from the existence of monopoles~\cite{mono} in
the $3$ dimensional Higgs model. The corresponding solitons of Higgs
models in arbitrary dimensions $d$ have been systematically shown to
exist, being constructed numerically in
\cite{dHiggs1,dHiggs2,dHiggs3}. The generalisation of the $d=3$ result of
\cite{JRb} to arbitary $d$ then follows almost trivially. Moreover, the
adaptation of the result of \cite{JRb} to the case of Goldstone model
backgrounds also follows systematically, by employing the solitons
presented in \cite{dGoldstone1,dGoldstone2,dGoldstone3}.

In section {\bf 2} we present the $4+d$ dimensional models on the space
with coordinates $x_M=(x_{\mu},x_m)$, $\mu=0,..,3$ labeling the Minkowski
space and $m=1,..2,d$ the codimension, and, we give the Ansatz separating
the varaibles $x_{\mu}$ and $x_m$ in the field equations. This describes
the dimensional descent. Both type ({\it i}) and type ({\it ii}) models,
i.e. with Dirac operators featuring both {\it partial} derivatives and
{\it covariant} derivatives, are presented in this section. The resulting
residual Dirac equations in $d$ dimensions will then be examined in detail
in the subsequent sections {\bf 3} and {\bf 4} respectively. In {\bf 3},
it will be shown that for models of type ({\it i}) the fermion cannot be
localised to the brane for any $d$. In section {\bf 4} type ({\it ii})
models will be analysed. In the first subsection of {\bf 4}, zero modes
of the residual Dirac equations in even $d\ge 4$ dimensional 'instanton'
backgrounds of YM systems will be constructed, resulting in the
{\em power localisation} of the fermion to the brane. In the second
subsection of {\bf 4}, the corresponding zero modes in all $d\ge 2$
dimensional 'soliton' backgrounds of Higgs (or their associated Goldstone)
systems will be constructed, resulting in the {\em exponential
localisation} of the fermion to the brane. A summary of the results is
given in section {\bf 5}, and three appendices have been supplied.
Appendix {\bf A} describes the (even) $d=2n$ dimensional YM models and
their 'instantons'. Appendix {\bf B} describes the $d$ dimensional Higgs
models and their 'solitons', and Appendix {\bf C} describes the Goldstone
counterparts of the latter.

\section{The model(s) and residual Dirac equations}
We will consider the following two types of fermionic actions, formally
expressed as
\bea
S_{\Psi}&=&
\int\ d^4x\ d^dx\ \left(\bar{\hat\Psi}\hat\Gamma^M\pa_M{\hat\Psi}
-\mu\,\si[\varphi]\ \bar{\hat\Psi}\hat\Psi\right)\label{act1}\\
S_{\Psi}&=&\int\ d^4x\ d^dx\ \left(\bar{\hat\Psi}\,
\hat\Gamma^MD_M{\hat\Psi}
-\mu\,\bar{\hat\Psi}\,\Xi[\varphi]{\hat\Psi}\right)\label{act2}
\eea
The first of these, \re{act1}, pertains to the type ({\it i}) family of
models featuring partial derivatives in the codimension. The components
of the spinor field on the codimension-$d$, in \re{act1},
are {\em isoscalar}. $\si[\varphi]$ in
\re{act1} is a scalar function of the fields $[\varphi]$ symbolising the
scalar and/or the YM field describing the brane Lagrangian in the
codimension-$d$. Specifically, it will be defined as the leading term
in the spherically symmetric restriction of the {\it topological current},
e.g. the Chern-Simons term in the case of YM. Both 'instanton' and
'soliton' backgrounds can be accommodated in this scheme.

The second, \re{act2}, represents type ({\it ii}) models in which
$D_M=(\pa_{\mu},D_m)$. The components of the spinor field on the
codimension-$d$, in \re{act2}, are {\em isospinor} except in the case
$d=2$ case when they are {\em isoscalar}. These models subdivide further
into two subclasses, ones with $\mu=0$, i.e. without a Yukawa term, and,
those with $\mu\neq 0$. The models \re{act2} with $\mu=0$ accommodate only
'instanton' backgrounds, while those with $\mu\neq 0$ accommodate only
'soliton' backgrounds, and in this case $\Xi[\varphi]$ is a matrix valued
function of $[\varphi]$.

The Dirac equations arising from \re{act1} and \re{act2} are, respectively
\bea
\hat\Gamma^M\pa_M\hat\Psi+\mu\ \si[\varphi_c]\,\hat\Psi&=&0
\label{dirac1}\\
\hat\Gamma^MD_M\hat\Psi+\mu\ \Xi[\varphi_c]\,\hat\Psi&=&0\,,\label{dirac2}
\eea
to be solved on the classical background $\varphi_c(r)$ to
be precised later. In the present work, we anticipate the use of radially
symmetric background solutions in terms of the radial variable
$r=\vert x_m\vert$ of the co-dimension, though a richer spectrum of
such backgrounds arises when this symmetry is relaxed.

Denoting the $4$-dimensional (spacetime) coordinates by $x_{\mu}$
and the coordinates of the codimension by $x_m$,
we represent the $4+d$ dimensional gamma matrices
$\hat\Gamma_M=(\hat\Gamma_{\mu},\hat\Gamma_m)$ by
\be
\label{gamma}
\hat\Gamma_{\mu}=\gamma_{\mu}\otimes\eins\qquad ,\qquad
\hat\Gamma_m=\gamma_5\otimes\Gamma_m
\ee
in terms of the $4$-dimensional gamma matrices $\gamma_{\mu}$ and their
chiral matrix $\gamma_5$, and the $d$-dimensional gamma matrices
$\Gamma_m$.

Our separability Ansatz, which also effects the dimensional descent, is
\be
\label{separable}
\hat\Psi(x_{\mu},x_m)=\Psi(x_{\mu})\otimes\psi(x_m)\ ,
\ee
and when applicable,
\be
\label{xi}
\Xi[\varphi(x_m)]=\eins\otimes\xi[\varphi(x_m)]\,.
\ee
In \re{xi}, the array $\xi[\varphi(x_m)]$ is a matrix valued array
whose size will be determined by the representaion in which the gauge
connection in the covariant derivative $D_m$ is. (In the generic case
this will be the Gamma matrix representation of $SO(d)$.) Detailed
definitions of $\xi$ for particular models to be considered, are given
in section {\bf 2.2}.

Using \re{gamma} and the separability Ansatz \re{separable},\re{xi}, the
Dirac equations \re{dirac1} and \re{dirac2} yield, respectively
\bea
\gamma^{\mu}\pa_{\mu}\Psi\otimes\psi+
\gamma^5\Psi\otimes\Gamma^m\pa_m\psi
+\mu\ \si_c\ \Psi\otimes\psi&=&0\label{ddirac1}\\
\gamma^{\mu}\pa_{\mu}\Psi\otimes\psi+
\gamma^5\Psi\otimes\Gamma^mD_m\psi
+\mu\ \Psi\otimes \xi_c\ \psi&=&0\,,\label{ddirac2}
\eea
in which we have used the notation $\si_c=\si[\varphi_c]$ and
$\xi_c=\xi[\varphi_c]$. If we now invoke the existence of the zero modes
of the Dirac field in 4 dimensional spacetime
\[
\gamma^{\mu}\pa_{\mu}\Psi=0\ ,
\]
and require that the Dirac spinor is chiral, i.e. that
\[
\gamma^5\Psi=\Psi\ ,
\]
then \re{ddirac1},\re{ddirac2} finally reduce to the {\em residual}
Dirac equations in $d$ dimensions
\bea
\left(\Gamma^m\pa_m+\mu\ \si_c\right)\psi&=&0\label{resdirac1}\\
\left(\Gamma^mD_m+\mu\ \xi_c\right)\psi&=&0\,.\label{resdirac2}
\eea
It is in order to mention at this point, that for the case of $\mu\neq 0$
models \re{act2}, the covariant derivative
$D_m$ in \re{resdirac2} will eventually be replaced by the partial
derivative $\pa_m$, since the detailed analysis of \re{resdirac2} to be
carried out subsequently is restricted only to Goldstone models associated
to the Higgs models described in Appendix {\bf C}, namely to the gauge
decoupled versions of the associated Higgs models described in the
Appendix {\bf B}. We stress that all our results are valid also for the
Higgs model 'soliton' backgrounds, and the only reason we eschew working
with the latter is that the corresponding analysis of \re{resdirac2}
yields qualitatively the same results as in the Goldstone case, and the
analysis in the latter case is somewhat simpler.

We will seek solutions to \re{resdirac1},\re{resdirac2} satisfying
\be
\label{s2}
\psi^{\prime}(0)<\infty ,\qquad ||\psi ||<\infty\,.
\ee
It will turn out that there exist no solutions of \re{resdirac1}
satisfying \re{s2} for any $d$, but will find such solutions to
\re{resdirac2} for all $d$.

\subsection{Definitions of $\si[\varphi]$}
$\si[\varphi]$ are defined as the leading terms in the {\it topological
currents} at infinity, i.e. when the fields defining them are spherically
symmetric. This subsection is subdivided in two parts.

In the first, we give the definition of $\si[\varphi]$ for a
topologically stable background supported by a $O(d+1)$ sigma model, as
well as a background supported by a $SO(d=2n)$ YM system in even, $d=2n$,
dimensions. The latter are the spherical--symmetrically restricted
Chern--Simons densities of $SO(d=2n)$ YM field, which turn out to be
described by essentially the same 'instanton' type profile
\be
-1\underset{r\leftarrow 0}\longleftarrow\si_c
\underset{r\rightarrow\infty}\longrightarrow 1\,,\label{inst}
\ee
as in the case of {\em even} dimensional sigma models.

In the second we consider backgrounds supported by Higgs or Goldstone
models, typified by 'soliton' profiles
\be
0\underset{r\leftarrow 0}\longleftarrow\si_c
\underset{r\rightarrow\infty}\longrightarrow 1\,,\label{soli}
\ee
as in the case of {\em odd} dimensional sigma models. Detailed
expositions of \re{inst} and \re{soli} are given in the following
two subsections.

\subsubsection{$\si[\varphi]$ for $O(d+1)$ sigma model and $SO(d=2n)$
pure YM backgrounds}

$d$ dimensional $O(d+1)$ sigma models and their topologically stable
solitons have been discussed extensively elsewhere~\cite{d+1} so we do
not elaborate on them here. Best known amongst these is the $d=2$
dimenesional {\it scale invariant} $O(3)$ sigma model whose solitons,
namely the well known Belavin-Polyakov vortices~\cite{BP}, are
evaluated in closed form. Here we are concerned only with the topological
boundary conditions the relevant solitons satisfy. Moreover, as noted
above, we will restrict to the case of radial (spherically symmetric)
solitons. So we state these, in terms of the $d+1$ component scalar
fields $\chi^a=(\chi^m,\chi^4)$, subject to the constraint
$\vert{\chi}^a\vert^2=1$:
\be
\label{ansatz1}
\chi^m=\sin f(r)\ \hat x^m\quad ,\quad
\chi^{d+1}=\cos f(r)
\ee
In \re{ansatz1} $\hat x^m=r^{-1}x^m$ is the unit vector
in the codimension. The topological charges stabilising the solitons of
these models are the winding numbers, which take on unit values provided
that the solutions satisfy the asymptotic conditions\footnote{The more
usual alternative $\lim_{r\to 0}f(r)=\pi\ ,\ \lim_{r\to\infty}f(r)=0$ is
not adopted, for the sake of making contact with
the usual asymptotics \re{YMboundary} for the corresponding YM fields.}
\be
\label{boundary}
\lim_{r\to 0}f(r)=0\quad ,\quad \lim_{r\to\infty}f(r)=\pi\ ,
\ee
Our definition of the function $\si_c=\si[f(r)]$ corresponding to the
solution $\varphi_c=f(r)$ of this model, is that given by \re{sigma}.

We list the functions $\si_c(d)$ for this model,
for the cases $d=2,\ d=4$ and $d=3,\ 5$, separately for
even and odd $d$. Up to unimportant multiplicative constant
depending on the angular volumes, we find following \cite{d+1},
\be
\label{even}
\si(d=2)\propto\ \cos f\quad ,\quad
\si(d=4)\propto\ \left(\cos f-\frac13\cos^3f\right)\,,\quad ...
\ee
for {\bf even} $d=2,4$ respectively, and
\be
\label{odd}
\si(d=3)\propto\ f-\frac12\sin 2f\quad , \quad
\si(d=5)\propto\ \left(\frac32 f-\sin 2f+\frac14\sin 4f\right)\,,\quad ...
\ee
for {\bf odd} $d=3,5$ respectively.
We see that the ranges of these topological charge densities are
quite different for even and odd $d$.

\re{even} and \re{odd} result, qualitatively, in the following profiles
of $\si_c$
\bea
+1\underset{r\leftarrow 0}\longleftarrow&\si_c&
\underset{r\rightarrow\infty}\longrightarrow -1
\qquad ,\qquad\mbox{\rm for even}\ \ d \label{boundaryeven}\\
0\underset{r\leftarrow 0}\longleftarrow&\si_c&
\underset{r\rightarrow\infty}\longrightarrow 1
\qquad ,\qquad\mbox{\rm for odd}\ \ \ d \ .\label{boundaryodd}
\eea

We next turn to the $SO(d)$ YM system in {\em even} $d$-diemensions,
for which the spherically symmetric Ansatz, analogous to \re{ansatz1}, is
\be
\label{YMansatz}
A_m=\frac{1-w(r)}{r}\,\Sigma_{mn}^{(\pm)}\hat x_n\quad ,\quad
\Sigma_{mn}^{(\pm)}=
-\frac14\left(\frac{1\pm\Gamma_{d+1}}{2}\right)[\Gamma_m,\Gamma_n]\,.
\ee
Analogously to \re{boundary}, the 'instanton' boundary conditions that
result in topologically stable (anti)-selfdual solutions to the systems
of YM hierarchies~\cite{YM}, are
\be
\label{YMboundary}
\lim_{r\to 0}w(r)=+1\quad ,\quad \lim_{r\to\infty}w(r)=-1\ ,
\ee
which coincides with \re{boundary} under the replacement
\be
\label{repl}
\cos f(r) \longleftrightarrow w(r)\,,\quad ...
\ee
Now in all {\em even} diemensions there exist Chern-Pontryagin charge
densities, whose spherically symmetric restrictions are the analogues
of \re{even}. Up to unimportant numerical factors, these densities in
dimensions $d=4$ and $d=6$ are
\be
\label{cp}
\si(d=4)\propto\ \left(w-\frac13w^3\right)\quad ,\quad
\si(d=6)\propto\ \left(w-\frac23w^3+\frac15w^5\right)\,.
\ee
Note that the first $(d=4)$ member of \re{cp} coincides with the
second $(d=4)$ member of \re{even} under the replacement \re{repl}.
This is a recurring coincidence. It is obvious now that the profile of
$\si_c$ in this case coincides with \re{boundaryeven}.

\subsubsection{$\si[\varphi]$ for $SO(d)$ Higgs/Goldstone model
backgrounds}
We now consider topologically stable backgrounds supported by
$d$ dimensional $SO(d)$ Higgs models~\cite{dHiggs1,dHiggs2,dHiggs3} and
their associated Goldstone
models~\cite{dGoldstone1,dGoldstone2,dGoldstone3}. 

The topologiocal charges of these
models are described by scalar fields $\phi^m$ ,
$m=1,2,...,d$, in $d$ dimensions. Apart from the various kinetic
terms, Goldstone models are distinguished by a symmetry breaking
self-interaction potential, leading to the important asymptotic
condition
\be
\label{inf-gold}
\lim_{r\to\infty}\vert\phi^m\vert^2=\eta^2
\ee
in which $\eta$ is the VEV with inverse dimensions of length. Here again,
we restrict to the radially symmetric fields
\be
\label{goldansatz}
\phi^m=\eta\,h(r)\,\hat x^m\,,
\ee
and for the special case of $d=2$ dimensions, the radially symmetric
{\em vorticity} $n$ field is
\be
\label{goldansatz2}
\phi^m=\eta\,h(r)\, n^m\quad ,\quad n^m=(\cos n\phi,\sin n\phi)\,.
\ee
The topological charges stabilising the solitons of these models are the
winding numbers, which take on the unit value for the following
asymptotic conditions
\be
\label{boundarygold}
\lim_{r\to 0}h(r)=0\quad ,\quad \lim_{r\to\infty}h(r)=1\ .
\ee
The function $\si_c$ is now expressed in terms of the classical soliton
profile $h(r)$. One difference from the sigma models of the previous
subsection however is, that the $d=1$ case for
Goldstone models, unlike the sigma models, does not trivialise but
coincides, for example, with the $\varphi^4$ model. Another difference
is that the winding number density
for the radial fields \re{goldansatz} does not take
qualitatively different expressions for even and odd $d$, as in
\re{even}-\re{odd}.

In both even and odd $d$-dimensional Higgs and Goldstone models, the
leading term in the winding number density turns out to be proportional to
\be
\label{varrho}
h(r)^d\qquad\Rightarrow\qquad\si_c\stackrel{def}=
\eta^{-d}\vert\phi^m\vert^d\equiv
\eta^{-d}\phi^d
\ee
and since in the following we will need only the asymptotic values and
not detailed behaviours of $\si_c(r)$, we omit the $d$-th power of
$h(r)$ in \re{varrho} and simply state the topologically meaningful
asymptotic behaviour
\be
\label{boundary-gold}
0\stackrel{r\leftarrow 0}\longleftarrow\si_c
\stackrel{r\rightarrow\infty}\longrightarrow 1
\qquad ,\qquad\mbox{\rm for all}\ \ \ d\,.
\ee

\subsection{Definitions of $\Xi[\varphi]$}
Unlike the quantities $\si[\varphi]$ presented in the previous subsection,
which are isoscalar, the quantities $\xi[\varphi]$ in \re{xi} are matrices
with isotopic indices. In models employing sigma model or YM 'instanton'
backgrounds in even codimension-$d$, $\mu=0$ so that $\Xi[\varphi]$ is
defined only for models employing 'soliton' backgrounds with $\mu\neq 0$.
With odd $d$ sigma model backgrounds in turn, there is no useful 
efinition for $\xi[\varphi]$. The reason is simple, and hinges on the
requirement that the residual Dirac equation \re{resdirac2}, like
\re{resdirac1}, should develop a mass term asymptotically in the
codimension.

Let us examine the Yukawa term in \re{resdirac2} in the asymptotic
region, which is subject to spherical symmetry. Consider the matrix
valued function $\xi[\varphi]$ in \re{xi} in terms of the two alternative
codimension fields $\chi^a=(\chi^m,\chi^{d+1})$, pertaining to the
$O(d+1)$ sigma model, and $\phi^m$, to the Higgs or Goldstone model. The
only natural forms for matrix valued $\xi$ in the gamma matrix
represenation in isospace are proportional to the
following matrix valued quantities
\be
\label{natural}
\xi\propto\Gamma_m\,\chi^m\quad ,\quad\xi\propto\Gamma_m\,\phi^m\,,
\ee
respectively. Inspecting the spherically symmetric Ans\"atze \re{ansatz1},
\re{goldansatz} and the asymptotics \re{boundary}, \re{boundarygold}
required, one sees that the only Yukawa term which leads to a
nonvanishing mass term is the second member of \re{natural}. Henceforth,
models of type \re{act2} with $\mu>0$ will be restricted to Higgs or
Goldstone model 'soliton' type backgrounds only, with the corresponding
Yukawa term determined by the second member of \re{natural}.
The type ({\it ii}) models considered are typified by the definitions
of the quantity $\xi$. There will be two such choices.

The first applies in the case where the gauge connection in the covariant
derivative is Abelian, which is the case only for $d=2$, e.g. in the
background of the usual Abelian Higgs model. In this case, it is possible
to take the Dirac field to be an {\it isoscalar}, and our choice for
$\xi$ is
\be
\label{xi2}
\xi=\si_1\si_m\phi^m=\phi^1\eins+i\phi^2\si_3\quad ,\ m=1,2\ ,
\ee
which is (Euclidean) Lorentz invariant.

The second concerns the case of generic codimension-$d$, where the
$SO(d)$ connection is non Abelian, our choice for $\xi$ is
\be
\label{xid}
\xi=\eins\otimes\Gamma_m\phi^m\,,
\ee
where the matrix $\eins$ is labeled by the spinor indices
and the matrix $\Gamma_m\phi^m$ is labeled by the isospinor indices.
While \re{xid} is defined for non Abelian backgrounds, i.e. for $d\ge 3$,
it applies also to the $d=2$ case formally. In that case, we express
the Abelian gauge connection, say $a_m$, in formally antihemitian form
\be
\label{abel}
A_m=\frac{i}{2}\ a_m\,\si_3
\ee
acting on the matrix valued Higgs field $\Phi=\phi^m\si_m$. The covariant
derivative in \re{resdirac2} for the $d=2$ case of the generic model is
then defined by the connection \re{abel}.

\section{Type ({\it i}) models with isoscalar
$\psi(_m)$}

We will show that type ({\it i}) models with action \re{act1}, on
backgrounds with either type of profile \re{inst} and \re{soli}
of the function $\si_c$, do not support solutions satisfying condition
\re{s2}, for any codimension $d$.
This section is divided into four subsections. In {\bf 3.1},
{\bf 3.2} and {\bf 3.3} we analyse the residual
Dirac equation \re{resdirac1} for $d=2$, $d=3$
and arbitrary $d$ respectively. This yields a set of coupled ordinary
differential equations, which are of the same form for all $d$. Then in
{\bf 3.4} we show that these equations do not have solutions satisfying
the required property \re{s2}.

\subsection{Codimension $d=2$}
The 2 component residual spinor $\psi(x_m)$ is subjeted to radial symmetry
\begin{eqnarray}
\label{s3}
\psi =\bigg(
\begin{array}{c}
f_1\ e^{im\phi}\\
f_2\ e^{im'\phi}
\end{array}
\bigg)\,,
\end{eqnarray}
with $m$ and $m'$, both integers. Denoting $m,m'$ instead by $l,l'$,
for uniformity of notation for all $d$, the variables $r$ and $\phi$ in
equation \re{resdirac1} separate for $l'=l+1$, resulting in
the pair of coupled first order equations
\begin{subequations}
\label{ss}
\begin{equation}
\label{ssa}
f^{\prime}_1-\frac{l}{r}f_1+\si_c\,f_2=0
\end{equation}
\begin{equation}
\label{ssb}
f^{\prime}_2+\frac{l+1}{r}f_2+\si_c\,f_1=0\,.
\end{equation}
\end{subequations}

\subsection{Codimension $d=3$}
The residual 2 component spinor $\psi(x_m)$ 
transforms as a spin-$\frac12$ spinor under 3 dimensional rotations,
and to achieve a separation of variables we employ the spinor
harmonics~\cite{VMK} $\Omega^{(\pm)}_{lm}$ to expand $\psi$
\be
\label{ansatz}
\psi=f_1(r)\,\Omega^{(+)}_{lm}\ +\ f_2(r)\,\Omega^{(-)}_{l'm}\,.
\ee
The spinor harmonics are defined as
\be
\label{spinor-h1}
\Omega^{(\pm)}_{lm}=C(l,1,l\pm 1)^{m-\frac12\,,+\,\frac12}_{\qquad\quad m}
\ Y_{l,\,m-\frac12}(\theta,\phi)\ \chi_{+\frac12}+
C(l,1,l\pm 1)^{m+\frac12\,,-\,\frac12}_{\qquad\quad m}
\ Y_{l,\,m+\frac12}(\theta,\phi)\ \chi_{-\frac12}
\ee
in which $Y_{l,\,m}(\theta,\phi)$ are the spherical harmonics and
$\chi_{\pm\frac12}$ are the constant valued $2$ component eigenvectors
for spin-$\frac12$
\be
\label{constspinors}
\chi_{+\frac12}=\left(\begin{array}{c}
1\\ 0 \end{array}\right)
\qquad ,\qquad \chi_{-\frac12}=\left(\begin{array}{c}
0\\ 1 \end{array}\right)\,.
\ee
Evaluating the Clebsch-Gordan coefficients in \re{spinor-h1} and
substituting \re{constspinors}, we have
\be
\label{spinor-h2}
\Omega^{(+)}_{lm}=\left(\begin{array}{c}
\sqrt{\frac{l+m+\frac12}{2l+1}}\,Y_{l,\,m-\frac12}(\theta,\phi) \\
\sqrt{\frac{l-m+\frac12}{2l+1}}\,Y_{l,\,m+\frac12}(\theta,\phi)
\end{array}\right)
\qquad ,\qquad
\Omega^{(-)}_{lm}=\left(\begin{array}{c}
-\sqrt{\frac{l-m+\frac12}{2l+1}}\,Y_{l,\,m+\frac12}(\theta,\phi) \\
\sqrt{\frac{l+m+\frac12}{2l+1}}\,Y_{l,\,m-\frac12}(\theta,\phi)\,.
\end{array}\right)
\ee
The result of acting with the gradient operator on the spherical
harmonics can be systematically calculated applying the Clebsch-Gordan
series~\cite{VMK}. Applying this to the residual Dirac equation
\re{resdirac1} with the Ansatz \re{ansatz}, and setting $l'=l+1$, we have
\be
\label{bigeqn}
\left(f^{\prime}_1-\frac{l}{r}\,f_1+\si_c\,f_2\right)\Omega^{(-)}_{l+1,m}+
\left(f^{\prime}_2+\frac{l+2}{r}\,f_2+\si_c\,f_1\right)\Omega^{(+)}_{l,m}
=0
\ee
resulting in

\begin{subequations}
\label{tt}
\begin{equation}
\label{tta}
f^{\prime}_1-\frac{l}{r}f_1=-\sigma_c f_2
\end{equation}
\begin{equation}
\label{ttb}
f^{\prime}_2+\frac{l+2}{r}f_2=-\sigma_c f_1\,.
\end{equation}
\end{subequations}
The similarity of \re{tta}-\re{ttb} with \re{ssa}-\re{ssb} is manifest
and holds for arbitrary codimension, as we shall see immediately below.

\subsection{Arbitrary codimension $d$}
To generalise to arbitrary codimension $d$, we need some preparation.
Let $p_k=-i\partial_k$, $\sigma_{ij}=i[\Gamma_i,\Gamma_j]/2$,  
$i,j=1,\ldots d$, and

$${\cal A}=\frac{1}{2}\sum_{i,j=1}^n \sigma_{ij}(x_ip_j-x_jp_i).$$
Squaring yields ${\cal A}^2 = (n-2){\cal A} + L^2$, where
$$L^2 = \sum_{i<j}(x_ip_j-x_jp_i)^2.$$
It is well known that the eigenvalues of $L^2$ are $l(l+d-2)$,
$l=0,1,\ldots.$  The corresponding possible eigenvalues of ${\cal A}$
are $-l$ and $l+d-2$, but by inspection only 0 for $l=0$. Thus the
list of ${\cal A}$ eigenvalues is
$0,-1,-2,\ldots$ plus $d-1,d,d+1\ldots$.
 
In analogy to the case $d=3$ we introduce the spinor harmonics 
$\Omega^{(+)}_{lM}$, $\Omega^{(-)}_{lM}$, where
$$L^2\Omega^{(\pm)}_{lM}=l(l+d-2)\Omega^{(\pm)}_{lM},$$
$${\cal A}\Omega^{(+)}_{lM}= -l\Omega^{(+)}_{lM}.$$
The index $M$ stands collectively for eigenvalues of angular momentum
operators which commute with ${\cal A}$.
Determining the spinor harmonics explicitly is more complicated than
for $d=3$, but not necessary for our purpose.

The operator $\Gamma^m\hat x_m$ transforms eigenspaces for $-l$ and
$(d-1+l)$ into each other, since
$${\cal A}(\Gamma^mx_m) = (\Gamma^mx_m)(d-1-{\cal A}),$$
as can be checked easily by multiplying the gamma matrices and using
obvious symmetries. Thus we can put
$$\Omega^{(-)}_{l'M}= \Gamma^m\hat x_m\,\Omega^{(+)}_{lM},$$
where $(d-2+l')=(d-1+l)$, thus $l'=l+1$ as for $d=2,3$.
Conversely we have
$$\Gamma^m\hat x_m\,\Omega^{(-)}_{l'M}= \Omega^{(+)}_{lM}.$$ 
Another multiplication of gamma matrices yields the Dirac operator
in the form
$$i\Gamma^mp_m =\Gamma^m\hat x_m\,\frac{d}{dr} +
\frac1r\,\Gamma^m\hat x_m\,{\cal A}.$$
With
$$\psi=f_1(r)\,\Omega^{(-)}_{l+1,M}\ +\ f_2(r)\,\Omega^{(+)}_{lM}$$
the Dirac equation takes the form
\begin{subequations}
\label{ttt}
\begin{equation}
\label{ttta}
f^{\prime}_1-\frac{l}{r}f_1+\sigma_c f_2=0
\end{equation}
\begin{equation}
\label{tttb}
f^{\prime}_2+\frac{l+d-1}{r}f_2+\sigma_c f_1=0\,.
\end{equation}
\end{subequations}
We see that both \re{ss} and \re{tt} for $d=2$ and $d=3$ are of the same
form as \re{ttt} for arbitrary $d$.

\subsection{Nonexistence}
The presentation here is adapted to both sigma model as well as to
Goldstone model backgrounds, namely for both 'inatanton', \re{inst},
and 'soliton', \re{soli}, profiles of $\si_c(r)$. Let us analyse
the $d=2$ equation \re{ss}, noting that the same conclusions hold for the
arbitrary case \re{ttt}.

In the $r\ll 1$ region, the solutions of equations (\ref{ssa},\ref{ssb})
which are differentiable at the origin, have the asymptotic forms
$$f_1\approx Ar^l\left(1+\frac{\sigma_c(0)^2}{4(l+1)}r^2\right)$$
$$f_2\approx-\frac{A\sigma_c(0)}{2(l+1)}r^{l+1}
\left(1+\frac{\sigma_c(0)^2}{4(l+2)}r^2\right)$$
In particular, $f_2(0)=0$ and $f_1(0)$ is finite for all possible values 
of $l$. At infinity, both $f_1$ and $f_2$ decay exponentially.

Now the equations (\ref{ss}) yield

$$-f_1f'_1+f_2f'_2 + \frac1r\left[lf_1^2+(l+1)f_2^2\right]=0.$$
Integrating over $r$ and using the said boundary conditions, one finds
$$\frac12\,f_1(0)^2+\int_0^\infty
\frac1r\left[lf_1^2+(l+1)f_2^2\right]\,dr =0,$$
which clearly is impossible.
There are therefore no solutions of \re{resdirac2} satisfying \re{s2}
in both the backrounds \re{inst} and \re{soli}, and in any $d$.

\section{Type ({\it ii}) models}

Type ({\it ii}) models with action \re{act2} separate in two main
subclasses, namely those with $\mu=0$, presented in the first
subsection {\bf 4.1}, and those with $\mu>0$,  presented in the
second subsection {\bf 4.2}. The quantity $\Xi[\varphi(x_m)]$, defined
in section {\bf 2.2}, will be specified further in two cases. The
first of these is a particular model with codimension-$d=2$
leading to the recovery of the result of \cite{LT,LTeco},
presented in subsection {\bf 4.2.1}, while the second is for the
generic codimension-$d\ge 2$ models, presented in subsection {\bf 4.2.2}.

$\mu=0$ models of type ({\it ii}) are defined exclusively in even
codimension-$d(=2n)$, since YM 'instanton' backgrounds exist only in even
dimensions~\footnote{It is also
possible for Grassmannian sigma models, in which case the Dirac operator
$D_m$ in \re{act2} features a {\it composite connection} in terms of the
Grassmannian field. Once we know this is a possibility, it is superfluous
to present it in detail here.}.
$\mu>0$ models of type ({\it ii}) are defined for all codimension-$d$,
in Higgs or Goldstone 'soliton' backgrounds. In the latter case we shall
eschew the Higgs backgrounds in favour of the corresponding {\it associated}
Goldstone backgrounds in which the gauge field is suppressed.

\subsection{Type ({\it ii}) models with $\mu=0$: even $d\ge 4$
isospinor $\psi(x_m)$}
Here, the solitons in whose background the residual Dirac equation
\re{resdirac2} is to be solved are restricted to the
'instantons' of $d=2n$ dimensional YM models described
in Appendix {\bf A}. To solve equation \re{resdirac2} with $\mu=0$ in
the spherically symmetric background for the YM connection \re{YMansatz},
we subject the isospinor $\psi(x_m)$ to spherical symmetry
\be
\label{sphspin}
\psi=f_1(r)\,\eins+f_2(r)\,\Sigma^{(\pm)}_m\,\hat x_m\,.
\ee
Substituting \re{sphspin} and the radial Ansatz \re{YMansatz} for
the YM connection into the residual Dirac equation \re{resdirac2},
with $\mu=0$, we find the familiar~\cite{JR4} solutions
\begin{subequations}
\label{zm}
\begin{equation}
\label{zma}
f_1=r^{-\frac12(d-1)}\e^{\frac12(d-1)\int\frac{w}{r}\,dr}
\end{equation}
\begin{equation}
\label{zmb}
f_2=r^{-\frac12(d-1)}\e^{-\frac12(d-1)\int\frac{w}{r}\,dr}\,.
\end{equation}
\end{subequations}
Adopting the asymptotics \re{YMboundary}, it is easy to see
that $f_1(r)$ satisfies the required condition \re{s2}, while $f_2(r)$
does not and must be rejected\footnote{When instead the alternative
asymptotics pointed out in footnote 2 is adopted, then
$f_2(r)$ satisfies \re{s2} and it is $f_1(r)$ that must be rejected.}.
It also follows from \re{YMboundary} that the relevant solution, e.g.
$f_1(r)$ has a {\em power decay} at infinity.

\subsection{Type ({\it ii}) models with $\mu\neq 0$}
In this case, the separability Ansatz \re{xi}, and the resulting
Yukawa terms like \re{natural}, will be specified in the two distinct
cases of a particular $d=2$ model featuring a column-valued Dirac 
spinor on the codimension, and the generic $d\ge 3$ models with
$SO(d)$ isospinor Dirac fields, which form square matrix arrays.
These cases will be presented in the following two subsections,
{\bf 4.2.1} and {\bf 4.2.2}.

In both {\bf 4.2.1} and {\bf 4.2.2}, the essential procedure is to so
select the separability Ansatz \re{xi}, such that the resulting
residual Dirac equation \re{resdirac2} turns out to be a known problem
leading to normaisable zero modes satisfying \re{s2}.

The analysis of the residual Dirac equation \re{resdirac2} for
these models will be restricted to 'soliton' backgrounds of the
{\it associated Goldstone} models, {\it cf.} Appendix {\bf C}, rather
than the corresponding backgrounds of the Higgs models, {\it cf.}
Appendix {\bf B}. The former are the gauge decoupled versions of the
latter and the results of these analyses are qualitatively the same.
Accordingly, \re{resdirac2} is effectively replaced in what follows by
\be
\left(\Gamma^m\pa_m+\mu\ \xi_c\right)\psi=0\,.\label{final2}
\ee
i.e. with $\pa_m$ replacing the $D_m$ in \re{resdirac2}.

\subsubsection{Type ({\it ii}) model with $\mu\neq 0\ :\ d=2$}
For $d=2$ the use of a doublet $(\phi^1,\phi^2)$ with a scalar and a
pseudoscalar component yields an interesting special situation. We
consider $U(1)$ gauge fields, such that in this model $\psi(x_m)$
does not carry an isotopic index. Specifying the separability
Ansatz \re{xi} with \re{xi2}, namely by
\[
\xi=\si_1\si_m\phi^m=\phi^1+i\si_3\phi^2\,,
\]
we end up essentially with the model of \cite{LT,JR}. The most interesting
feature of this model is the presence of fermionic zero modes for Abelian
backgrounds with any vorticity $n$.

With \re{xi2} in the residual Dirac equation \re{resdirac2}, one proceeds
to solve the latter in the background of the $d=2$ Higgs 'soliton',
namely the usual Nielsen-Oleson vortex or another ($p\ge 2$) member of the
hierarchy in Appendix {\bf B}, e.g. the vortex of the system \re{11}.

Alternatively, as in effect we will, one can solve \re{resdirac2}
in the background of the $d=2$ Goldstone 'soliton' of the $p=2$ member
of the hierarchy in Appendix {\bf C}, namely the vortex of the system
\re{11g}, i.e. the one resulting from the gauge decoupling of the system
\re{11}. We restrict the subsequent analysis to that of \re{resdirac2}
in the associated Goldstone 'soliton' background.

Substituting \re{xi2} with $\phi^m$ given by \re{goldansatz2} and the
radially symmetric Ansatz \re{s3} for $\psi$, in \re{resdirac2} for $d=2$,
the latter separates for $m'=m+1$,
and reduces to the pair of coupled first order equations
\begin{subequations}
\label{ss2}
\begin{equation}
\label{ssa2}
f^{\prime}_1-\frac{m}{r}f_1+\eta\,h\,f_2=0
\end{equation}
\begin{equation}
\label{ssb2}
f^{\prime}_2+\frac{n+m+1}{r}f_2+\eta\,h\,f_1=0\,.
\end{equation}
\end{subequations}
The Dirac equations in \cite{JR,LT,RjS} reduce to Eqns.
\re{ssa2},\re{ssb2}, reproducing the $d=2$ result of \cite{LT,RjS},
for completeness.

\subsubsection{Type ({\it ii}) models with $\mu\neq 0\ :\ d\ge 2$
isospinor $\psi(x_m)$}
The situation here is similar to the case of 'instanton' backgrounds
considered in {\bf 4.1} and likewise the isospinor Dirac field subject
to spherical symmetry is
\be
\label{NAfermi}
\psi=f_1(r)\,\eins+f_2(r)\,\Gamma_m\hat x_m\,.
\ee
Note here that in \re{NAfermi} we have $\Gamma_m$ in all $d$ dimensions,
while in \re{sphspin} we have chiral matrices $\Sigma_m^{(\pm)}$ in
$d=2n$, even, dimensions.

The separability Ansatz in these cases is specified by \re{xid}, namely
\[
\xi=\eins\otimes\Gamma_m\phi^m\,.
\]
The residual Dirac equation \re{final2} in the Goldstone 'soliton'
background now separates\footnote{Note here that for the case $d=2$,
where the Abelian Higgs (or Goldstone) background \re{goldansatz2}
is radial for all vorticity $n$, this separation can take place only
for the {\it unit} vorticity $n=1$ background. This contrasts with the
model of \cite{JR,LT} presented above in {\bf 4.2.1}.} and yields the
following pair of first order equations

\begin{subequations}
\label{rr}
\begin{equation}
\label{rra}
f^{\prime}_1+\eta h\,f_1=0
\end{equation}
\begin{equation}
\label{rrb}
f^{\prime}_2+\frac{d-1}{r}f_2+\eta h\,f_2=0\,,
\end{equation}
\end{subequations}
leading to
\begin{subequations}
\label{uu}
\begin{equation}
\label{uua}
f_1=\e^{-\eta\int h\,dr}
\end{equation}
\begin{equation}
\label{uub}
f_2=\frac{1}{r^{d-1}}\e^{-\eta\int h\,dr}\,.
\end{equation}
\end{subequations}
Given the asymptotics of $h(r)$, \re{boundarygold}, and the behaviour
of $h(r)$ near the origin to be
\be
\label{origin}
h(r)\approx b\,r\,,
\ee
it follws that both $f_1$ and $f_2$ vanish asymptotically in the $r\gg 1$
region as required, but only $f_1$ converges in the $r\ll 1$ region while
$f_2$ diverges and must be rejected. The result is one normalisable zero
mode, $f_1(r)$. It follows from \re{boundarygold} that the solution
$f_1(r)$ has an {\em exponential decay} at infinity.

The corresponding result, for \re{resdirac2} in the Higgs 'soliton'
background, can be readily found using the Ansatz \re{YMansatz}, with
the matrices $\Sigma_m^{(\pm)}$ there replaced by $\Gamma_m$,
{\it viz.} \re{YMansatzG} in Appedix {\bf B}. The result is qualitatively
the same, with the zero mode $f_1(r)$ still localised exponentially,
except that the energy density of the YMH
background brane is power localised rather than that of the Goldstone
background brane analysed here, which is exponentially localised.

Instead of one such zero mode, it is possible to construct a family of
such solutions by relaxing the constraint of spherical symmetry. In
particular, imposing only axial symmetry characterised by a vortex
number $n$, a family of such solutions
labeled by $n$ can be found. We do not present the details here.

\section{Summary}
We have addressed the problem of extending the mechanism of confining
a fermion to the brane in a $4+1$ dimensional model, proposed in
\cite{RS}, to the case of $4+d$ dimensional models, for arbitrary $d$.

In the model of \cite{RS,R}, the confinement mechanism relies on the
fact that a scalar field model in the $1$ dimensional extra
coordinate, i.e. the codimension-$1$, supports topologically stable
'soliton' solutions. This scalar field enters the $5$ dimensional
fermionic model through a Yukawa interaction term and results in the
Dirac equation of the system developing a mass term asymptotically in
the codimension, which is responsible for the confinement.

To extend this mechanism to higher dimensions, it seems~\cite{R} natural
to employ some field theoretic model on the codimension-$d>1$ that
supports topologically stable finite energy solutions. We considered
candidates for such models which support either 'soliton' or 'instanton'
like solutions. Our nomenclature throughout was that 'solitons' are
supported by Higgs or Goldstone models, most notably featuring
dimensionful scalar fields whose self-interaction potential leads to
symmetry breaking. 'Instantons' on the other hand are supported by
purely YM models or by sigma models, in even dimensions.

We proposed two types of models, in both of which the separation of
the Minkowski space coordinates $x_{\mu}$ from the codimension-$d$
coordinates $x_m$ was effected by an Ansatz, which also resulted in
the dimensional descent of the Dirac equation in $4+d$ dimensions,
to one in $d$ dimensions, which we referred to as the {\it residual
Dirac equation}. The solutions of the latter were what described the
localisation of the fermion to the brane.

The first type, ({\it i}), of models was characterised by a Dirac
operator, which featured a partial derivative in all components of the
differential operator. Consequently, the
residual Dirac spinors were {\em isoscalars}. The information on the
topologically stable solutions on whose background the residual Dirac
equation was solved, was encoded in a scalar coefficient in the Yukawa
term. This quantity was a descendent of the topological invariant of
the background system. It was found that the residual Dirac equation of
these models did not support normalisable zero modes.

The second type, ({\it ii}), of models was characterised by a Dirac
operator, which featured a partial derivative in the Minkowskian
components of the differential operator and a covariant derivative for
the components on the codimension. Consequently, the
residual Dirac spinors were {\em isospinors} for $d\ge 3$ when the gauge
group was non Abelian, and only in the $d=2$ case when the gauge group
was Abelian it was {\em isoscalar}. Type ({\it ii}) models did result
in normalisable zero modes for the residual Dirac equations, provided that
the Yukawa term was chosen appropriately, and in the case of pure YM
'instanton' backgrounds this meant its absence. When 'instanton'
backgrounds were employed localisation to the brane featured a power
of $r$, while employing 'soliton' backgrounds of Higgs or Goldstone
models resulted in exponential localisation.

A nontrivial aspect of our results is that in the models presented
in sections {\bf 4.2}, the Higgs models, i.e. the YMH systems,
can be contracted down to the {\it associated} Goldstone models by
elimination of the gauge fields, {\it cf.} Appendix {\bf C}. Whether this
additional feature in constructing such models is of any physical
advantage is not obvious, but we note that the energy density of the
Higgs model brane system is power localised, while that in the
Goldstone case is exponentially localised. Also, in some of the work of
\cite{LT}, a Goldstone model has been employed, albeit a model with
divergent energy.

\bigskip
\noindent
{\bf\large Acknowledgements}

\noindent
One of us (D.H.Tch.) thanks the Alexander von Humboldt Foundation for an
invitation to Bonn where this work was started. This work is carried out
in the framework of Enterprise--Ireland Basic Science Research Project
SC/2003/390.

\appendix
\section{Yang--Mills models in $d=2n$ dimensions}
\setcounter{equation}{0}
\renewcommand{\theequation}{A.\arabic{equation}}

In all even $d=2n$ dimensions, it is possible to construct Yang--Mills
(YM) models which support 'instanton' like solutions, which are
topologically stable and their energy integrals are finite. The
fundamental relation that ensures the existence of 'instanton' in this
hierarchy~\cite{YM} of YM systems are the inequalities stating the
lower bounds on the energy integrals, given by the appropriate
topological charges, namely the Chern-Pontryagin (C-P) numbers.

Using the convenient notation for the $2p$-form $p$-fold totally
antisymmetrised product of the curvature $2$-form $F\equiv F(2)$,
\be
\label{not}
F(2p)_{\mu_1\mu_2...\mu_{2p}}=F_{[\mu_1\mu_2}F_{\mu_3\mu_3}...
F_{\mu_{2p-1}\mu_{2p}]}\,,\quad {\rm totally\ antisymmetrised\ in}\ 
[\mu_1\mu_2...\mu_{2p}],
\ee
the simplest such inequality for a $2(p+q)$ dimensional YM system states
\be
\label{ineq}
\mbox{Tr}\left(\vert F(2q)\vert^2+
\ka^{2(p-q)}\frac{(2q)!}{(2p)!}\vert F(2p)\vert^2\right)\ge2\ka^{p-q}
\mbox{Tr}\,F\wedge F\wedge ...\wedge F\,,\quad p+q\ \ {\rm times}
\ee
where $\ka$ is a constant with the dimensions of length (if $p>q$). The
left hand side defines the energy density of the YM system and the
right hand side is proportinal to the $(p+q)$-th C-P density.

In the special case where $p=q$, the $4p$ dimensional YM systems are
scale invariant and the inequality \re{ineq} can be saturated. If the
gauge group is chosen to be $SO(4p)$ and the gauge fields are in the
chiral representations thereof then instanton solutions can be
evaluated explicitly in the spherically symmetric cases, but these do not
interest us here.

What is relevant in the present context is the family of YM models in
$4p$ dimensions from which the $d(<4p)$ dimensional Higgs models described
in Appendix {\bf B} below are constructed. Also relevant are the
$2(p+q)=2n$ dimensional models defined by \re{ineq}, in any even
dimension, whose solutions~\cite{BT} provide the 'instanton' backgrounds
used in section {\bf 4.1}. Solutions to both these types of YM models,
scale invariant or otherwise, are 'instantons' in the sense that at infinity
the gauge connection is {\it pure gauge} satisfying
\be
\label{pg}
A_m\underset{r\to\infty}\longrightarrow g^{-1}\pa_mg\,,
\ee
such that in the spherically symmetric case \re{YMansatz} the 'instanton'
profile \re{YMboundary} obtains.

\section{Higgs models in $d$ dimensions}
\setcounter{equation}{0}
\renewcommand{\theequation}{B.\arabic{equation}}

In this Appendix, we describe the Higgs
models~\cite{dHiggs1,dHiggs2,dHiggs3} in $d$ dimensions which support
finite energy topologically invariant soliton solutions.

In any given dimension $d$, a hierarchy of Higgs models supporting
solitons can be systematically constructed by subjecting the $p$-th
member of the Yang--Mills hierarchy~\cite{YM}, {\it cf} Appendix
{\bf A}, in dimension $4p>d$, to
dimensional reduction down to $d$ dimensions. The descent mechanism
essentially consists of the imposition of a symmetry, which results in
the breaking of the gauge group of the original $4p$ dimensional YM
system, and at the same time the components of the gauge connection on the
extra $4p-d$ dimensions appear as Higgs fields in the residual $d$
dimensional system, namely in a Higgs model. By choosing the gauge group
of the $4p$ dimensional system suitably, the gauge group of the $4p$
dimensional YM system breaks down to $SO(d)$, yielding the required $d$
dimensional $SO(d)$ Higgs model~\cite{dHiggs1,dHiggs2,dHiggs3}. For
simplicity we will restrict to the scale invariant YM systems for
the present purpose.

Now the action density of the $4p$ dimensional scale invariant YM system
is bounded from below by the $2p$-th Chern--Pontryagin (C-P) density. It
turns out that under the descent mechanism described in the previous
paragraph, this topological lower bound translates to a new lower bound on
the energy density of the residual $d$ dimensional Higgs model. The lower
bound is given by the residual C-P density, which now depends on the
residual gauge group, not necessarily $SO(d)$. Such lower
bounds might be described as {\it Bogomol'nyi bounds}. That this
residual C-P density is a topological charge density follows from the
fact that it can be shown to be a total divergence\cite{totdiv}, whose
resulting surface integral turns out to be finite subject to the usual
{\em symmetry breaking} asymptotics of the Higgs field {\em provided}
that the residual gauge connection also exhibits the requisite
asymptotics.

In the particular case of interest, namely when the residual gauge group
is arranged to be $SO(d)$, the Higgs multiplet is
\be
\label{vecH}
\Phi=\Gamma_m\,\phi^m\,,
\ee
where the index $m=1,2,..,d$ labels also the coordinates $x_m$, in the
same notation as above. The {\em symmetry breaking} condition of the
Higgs field can then be stated as 
\be
\label{symbr}
|\phi|^2=\phi^m\phi^m\underset{r\to\infty}\longrightarrow\eta^2\,,
\ee
where $\eta$ is the VEV, related to the compactification scale used in
the dimensional descent from $4p$ dimensions. For $d\ge 3$, i.e. when the
residual gauge group is non Abelian, the Higgs field points along the
unit vector $\hat x_m$ in the $r\gg 1$ asymptotic region, i.e. on the
$d-1$ sphere. In this region the gauge group breaks down to $SO(d-1)$
and the connection field decays as $r^{-1}$. In the Dirac gauge, where
the Higgs field in the $r\gg 1$ asymptotic region points along the
$d$-th direction, the connection develops a
semi-infinite line singularity in the $x_d$ direction, which is an
artefact of this gauge. This analogy with the familiar case of the
monopole~\cite{mono} in the $d=3$ case is complete and the residual
$SO(d)$ connection behaves as
\be
\label{halfpg}
A_m\underset{r\to\infty}\longrightarrow\frac12\,g^{-1}\pa_mg\,,
\ee
namely as {\em half a pure-gauge}, rather than as
one {\em pure-gauge} like an instanton. It is for this reason that  above,
we have called the finite energy topologically stable solutions
of Higgs models 'solitons', in contrast with the corresponding
solutions of even dimensional sigma models and YM systems, as 'instantons'.

In terms of the spehrically symmetric Ansatz, which is \re{YMansatz}
with $\Sigma_{mn}^{(\pm)}$ now replaced by $\Gamma_{mn}$
\be
\label{YMansatzG}
A_m=\frac{1-w(r)}{r}\,\Gamma_{mn}\hat x_n\,,
\ee
the asyptotics of the function $w(r)$ corresponding to \re{halfpg} are
\be
\label{YMsoliboundary}
\lim_{r\to 0}w(r)=+1\quad ,\quad \lim_{r\to\infty}w(r)=0\,.
\ee
The only exception to the property \re{halfpg} is the $d=2$ Higgs model,
in which case the boundary of the space is not sufficiently large to
accommodate a Dirac gauge.

The most familiar Higgs models which can be construced in this scheme
descend from the usual $SU(2)$ YM system in $4$ dimensions, i.e. from the
first ($p=1$) member of the YM hierarchy~\cite{YM}, down to $d=3$ and
$d=2$ respectively. In $d=3$ one finds the $SO(3)$ Georgi-Glashow model in
the Prasad-Sommerfield limit, and in $d=2$, the familiar Abelian Higgs
model. To illustrate this scheme further we have to proceed to $p\ge 2$,
and for the sake of ease of presentation, we restict to the first two
nontrivial examples. These are the $SO(d)$ Higgs models arising from
$(p=2,d=3)$ and $(p=2,d=2)$.

The $d=3$, $SO(3)$ Higgs model~\cite{dHiggs2} is defined by the
Lagrangian $\cal L$ which is bounded from below by the topological
charge density $\varrho$,
\begin{eqnarray}
{\cal L}
& = & \mbox {Tr} \bigg( \{F_{[ij},D_{k]} \Phi \}^2
+6\lambda_3 (\{ S,F_{ij} \} +[D_i \Phi ,D_j \Phi])^2 \: , \nonumber \\
& & \qquad \qquad \qquad \qquad +27\lambda_2 \{ S,D_i \Phi \}^2
+54 \lambda_1 S^4 \bigg)
\label{10} \\
\varrho
& = &
36\vep_{ijk} \pa_k \mbox {Tr} \bigg[ \phi (3\eta^4 -2\eta^2 \Phi^2
+\frac35\Phi^4)F_{ij} \nonumber \\
& & \qquad \qquad \qquad -2\eta^2 \Phi D_i \Phi D_j
\Phi -\frac25\Phi^2 (2\Phi D_i \Phi -D_i \Phi \Phi)D_j \Phi \bigg],
\label{12}
\end{eqnarray}
in which $\Phi$ is given by \re{vecH}, $S=\eta^2-\Phi^2$, and
the manifestly total divergence
form of $\varrho$ is displayed in \re{12}. Solutions
to this system, were constructed in \cite{dHiggs2}.

The $d=2$, $SO(2)$ or $U(1)$ Higgs model~\cite{dHiggs3} is defined by the
Lagrangian $\cal L$ which is bounded from below by the topological charge
density $\varrho$,
\bea
{\cal L}&=&
\la_2[(\eta^2-|\varphi|^2)F_{ij}+iD_{[i}\varphi^*D_{j]}\varphi]^2+
24\la_1(\eta^2-|\varphi|^2)^2|D_i\varphi|^2+18\la_0(\eta^2-|\varphi|^2)^4
\label{11}\\
\varrho&=&\vep_{ij}\,\pa_i\left[\eta^6A_j-3i\left(\eta^4-\eta^2|\varphi|^2
+\frac13(|\varphi|^2)^2\right)\varphi D_j\varphi^*\right]\label{13}
\eea
where we have used the complex valued Higgs field $\varphi=\phi^1+i\phi^2$,
and again the topological density $\varrho$ is displayed in manifestly
total divergence form.

It is easy to see that the leading terms making a nonvanishing
contribution to the integrals of the topological charge densities
\re{12} and \re{13}, respectively, are the {\em magnetic charge}
of the monopole and the {\em winding number} of the Nielsen--Oleson
vortex.

\section{Goldstone models associated to the Higgs models in
$d$ dimensions}
\setcounter{equation}{0}
\renewcommand{\theequation}{C.\arabic{equation}}
In this Appendix we define what we have referred to in the above as
the Goldstone models
{\it associated}~\cite{dGoldstone1,dGoldstone2,dGoldstone3} to the Higgs
models in $d$ dimensions described in Appendix {\bf B} above,
which also support soliton solutions.

The aspect of \re{10} and \re{11} that concerns us here is that
both possess gauge decoupling limits, which is a consequence of the fact
that when the gauge fields are removed~\cite{dGoldstone1} from these
densities, the remaining density still satisfies the Derrick scaling
requirement. Indeed, in \cite{dHiggs3} this gauge decoupling was
demonstrated concretely for the numerically constructed solutions:
specifically, the Goldstone 'soliton' in \cite{dGoldstone3} is obtained
by gauge decoupling the 'soliton' in \cite{dHiggs3}. This
feature is in stark contrast with the same procedure for the {\em usual}
$d=3$ and $d=2$ Higgs models arrived at from $p=1$ YM, i.e. the
Georgi-Glashow model and the Abelian Higgs model, in which
cases the solitons do not persist after gauge decoupling.

It is thus possible~\cite{dGoldstone1} to find Goldstone models
associated to each $d$ dimensional $SO(d)$ Higgs model descended from all
$p\ge 2$ members of the YM hierarchy. We demonstrate this prescription in
the two examples considered explicitly in Appendix {\bf B}.

In the $(p=2,d=3)$ model, eliminating the gauge field in \re{10},\re{12}
we find the Lagrangian and topological charge density of the
{\it associated} $(p=2,d=3)$ Goldstone model
\begin{eqnarray}
{\cal L}
& = & \mbox {Tr} \bigg(6\lambda_3 [\pa_i \Phi ,\pa_j \Phi]^2
+27\lambda_2\eta^4(\eta^2-\Phi^2)^2\pa_i\Phi^2+
54 \lambda_1(\eta^2-\Phi^2)^4 \bigg)
\label{10g} \\
\varrho
& = &
36\vep_{ijk} \pa_k \mbox {Tr} \bigg[-2\eta^2 \Phi \pa_i \Phi \pa_j
\Phi-\frac25\Phi^2(2\Phi\pa_i\Phi -\pa_i \Phi \Phi)\pa_j \Phi \bigg]\,.
\label{12g}
\end{eqnarray}
Note that covariant derivatives in \re{10} are replaced by partial
derivatives in \re{10g}. This is the rationale behind the corresponding
replacement of $D_m$ in \re{resdirac2} by $\pa_m$ in \re{final2}.

The system described by \re{10g} is related to that considered in
\cite{dGoldstone2}, except that in the latter, we have selected
specific values of the couplings $\lambda_{(i)}$, and also
replaced the symmetry breaking potential in \re{10g} by other postitive
definite (and symmetry breaking) potentials, without compromising the
existence of the solutions. Note that in the asymptotic region, \re{12g}
is equivalent to the {\it winding number} density.

In the $(p=2,d=2)$ model, eliminating the gauge field in \re{11},\re{13}
we find the Lagrangian and topological charge density of the
{\it associated} $(p=2,d=2)$ Goldstone model
\bea
{\cal L}&=&
\la_2|\pa_{[i}\varphi^*\pa_{j]}\varphi|^2+
24\la_1(\eta^2-|\varphi|^2)^2|\pa_i\varphi|^2+
18\la_0(\eta^2-|\varphi|^2)^4
\label{11g}\\
\varrho&=&-3i\vep_{ij}\,\pa_j\left[\left(\eta^4-\eta^2|\varphi|^2
+\frac13(|\varphi|^2)^2\right)\varphi\pa_j\varphi^*\right]\,,\label{13g}
\eea
which is the system investigated in \cite{dGoldstone3}, except that in the
latter two specific symmetry breaking potentials in addition to that
in \re{11g} were employed in the numerical construction of the solutions.

The above described procedure of constructing the {\it associated}
Goldstone model for any $SO(d)$ Higgs model characterised by any
$(p,d)$, can be carried out.

\begin{small}

\end{small}

\end{document}